\documentclass[showkeys,
reprint,
superscriptaddress,
 amsmath,amssymb,
 aps,
pra,
floatfix,
]{revtex4-2}

\usepackage{svg}
\usepackage{graphicx}
\usepackage{dcolumn}
\usepackage{bm}
\usepackage[]{xcolor}
\usepackage{subcaption}
\usepackage{physics}
\usepackage{qcircuit}
\usepackage{appendix}

\begin{document}


\title{Quantum circuit synthesis for fermionic excitations in coupled cluster theory}

\author{Yu-Hao Chen}%
\email{d13222001@ntu.edu.tw}
\affiliation{Department of Physics, National Taiwan University, Taipei, Taiwan ROC}%
\affiliation{Department of Artificial Intelligence, Chang Gung University, Taoyuan, Taiwan ROC}%
\author{Renata Wong}%
\thanks{Corresponding author}%
\email{renata.wong@cgu.edu.tw}
\affiliation{Department of Artificial Intelligence, Chang Gung University, Taoyuan, Taiwan ROC}%
\affiliation{Department of Neurology, Chang Gung Memorial Hospital, Keelung, Taiwan ROC}%
\affiliation{Artificial Intelligence Research Center, Chang Gung University, Taoyuan, Taiwan ROC}




\begin{abstract}
This work provides a step-by-step quantum computing derivation of the unitary coupled cluster singles and doubles ansatz. We show how the structure of the ansatz emerges from fermionic algebra under unitary constraints. By explicitly connecting second quantization, Jordan–Wigner mapping, and quantum circuit synthesis, we clarify conceptual gaps between quantum physics and quantum computing implementations, particularly regarding operator locality, commutation structure, and hardware realization. We exemplify the procedure on the singles and doubles coupled cluster ansatz for the hydrogen molecule. 

\end{abstract}

\maketitle

\section{\label{sec:intro}Introduction}

Quantum computing offers a potential approach to simulating quantum many-body systems, including problems in electronic structure. One widely studied method is the variational quantum eigensolver (VQE) \cite{cerezo}, which uses parameterized quantum circuits (ansätze) together with classical optimization to estimate molecular ground-state energies. The choice of ansatz plays an important role in VQE, particularly in quantum chemistry. Among the commonly used choices, the unitary coupled cluster singles and doubles (UCCSD) ansatz is popular because it is based on established methods in electronic structure theory.

The UCCSD ansatz is based on classical coupled cluster (CC) theory, which describes electron correlation using an exponential operator built from excitation operators. This approach provides a size-extensive description of correlated wavefunctions. Adapting CC theory for quantum computing, however, involves several challenges. Fermionic operators must first be mapped to qubit operators, the non-unitary CC operator must be replaced by a unitary form, and the resulting expressions must be translated into quantum circuits that can run on actual hardware.

Previous studies (see e.g. \cite{abhinav}) have approached this problem from different perspectives. Work in quantum chemistry typically starts from second quantization and reformulates the theory for quantum computers, while quantum algorithm studies tend to focus on circuit design, efficiency, and optimization. The steps linking fermionic many-body theory to quantum circuits are therefore not always discussed in a unified way. For example, the Jordan–Wigner transformation is often presented as a tool for converting fermionic operators into qubit operators, with less emphasis on how it preserves fermionic anticommutation relations at the operator level.

In this work, we derive the UCCSD ansatz and its quantum circuit implementation step by step, starting from the principles of quantum computation. Instead of adapting classical coupled cluster theory to a quantum computer, we begin with unitary state evolution and show how the ansatz can be constructed from this requirement. We then explain how fermionic excitation operators are represented by qubit operators and how these operators lead to a practical quantum circuit implementation.

The Jordan–Wigner transformation plays a key role in this derivation by mapping fermionic operators to sequences of Pauli operators. Among the available fermion-to-qubit mappings, we use Jordan–Wigner because its construction is intuitive and makes the connection between fermionic operators and qubit operations easier to follow. We examine how the mapping preserves fermionic anticommutation relations through parity strings that account for the occupations of other orbitals. Using the hydrogen molecule in a minimal basis as an example, we derive the Pauli operator expressions for single and double excitations and show how they can be implemented as quantum circuits.

We also discuss practical issues in implementing the ansatz as a quantum circuit, including basis changes, parity encoding with entangling gates, and rotation gates supported by the target hardware. Since excitation operators generally do not commute, the unitary coupled cluster operator must be approximated using Trotterization, and the order of the excitation operators can affect the resulting circuit. These details show that UCCSD does not have a single circuit implementation. Different choices in circuit construction may affect both the states the ansatz can represent and the performance of VQE optimization.

By working through the steps from second quantization to qubit operators and quantum circuits, this work connects the language of quantum chemistry with its implementation on quantum processors. The derivation shows how many-body wavefunction ansätze are converted into executable quantum algorithms, providing a clearer understanding of the underlying physics and practical guidance for implementing quantum simulations.

\section{Quantum states and quantum gates}

To make this work self-contained, we briefly introduce the basic concepts of quantum computation used in this paper.

\subsection{Quantum states}

A pure state of a single-qubit system is represented by a normalized state vector
\[
|\psi\rangle = \alpha |0\rangle + \beta |1\rangle
=
\begin{bmatrix}
\alpha\\
\beta
\end{bmatrix},
\qquad
|\alpha|^2+|\beta|^2=1,
\]
where \(\alpha,\beta\in\mathbb{C}\) are probability amplitudes, and
\[
|0\rangle=\begin{bmatrix}1\\0\end{bmatrix},
\qquad
|1\rangle=\begin{bmatrix}0\\1\end{bmatrix}
\]
are the computational basis states.

The complex conjugate transpose of a quantum state $|\psi\rangle$ is given as
\[
\langle \psi | = |\psi\rangle^\dagger.
\]

For a system of \(n\) qubits, the state belongs to a \(2^n\)-dimensional Hilbert space and can be expressed as
\[
|\Psi\rangle =
\sum_{x\in\{0,1\}^n} c_x |x\rangle,
\qquad
\sum_{x\in\{0,1\}^n}|c_x|^2=1,
\]
where \(|x\rangle=|x_1x_2\cdots x_n\rangle\) denotes a computational basis state and \(c_x\in\mathbb{C}\) is its probability amplitude. The computational basis states are tensor products of single-qubit basis states; for example,
\[
|01\rangle=|0\rangle\otimes|1\rangle
=\begin{bmatrix}0\\1\\0\\0\end{bmatrix},
\]
using the basis ordering \(\{|00\rangle,|01\rangle,|10\rangle,|11\rangle\}\).

A multi-qubit state is not necessarily a product of individual qubit states. States that cannot be expressed as tensor products of single-qubit states are entangled. For example,
\[
|\Phi^+\rangle=\frac{|00\rangle+|11\rangle}{\sqrt{2}}
\]
is an entangled two-qubit state.

\subsection{Quantum gates}

The Pauli operators are defined as
\[
X=\begin{bmatrix}0&1\\1&0\end{bmatrix},
\qquad
Y=\begin{bmatrix}0&-i\\i&0\end{bmatrix},
\qquad
Z=\begin{bmatrix}1&0\\0&-1\end{bmatrix}.
\]
They are Hermitian and unitary, and satisfy \(X^2=Y^2=Z^2=I\).

The Hadamard gate is
\[
H=\frac{1}{\sqrt{2}}
\begin{bmatrix}1&1\\1&-1\end{bmatrix},
\]
which transforms the computational basis states into equal superpositions:
\[
H|0\rangle=\frac{|0\rangle+|1\rangle}{\sqrt{2}},
\qquad
H|1\rangle=\frac{|0\rangle-|1\rangle}{\sqrt{2}}.
\]

Rotations about the three coordinate axes are defined by
\[
R_X(\theta)=e^{-i\theta X/2},\qquad
R_Y(\theta)=e^{-i\theta Y/2},\qquad
R_Z(\theta)=e^{-i\theta Z/2}.
\]
Their matrix representations are
\[
R_X(\theta)=
\begin{bmatrix}
\cos(\theta/2)&-i\sin(\theta/2)\\
-i\sin(\theta/2)&\cos(\theta/2)
\end{bmatrix},
\]
\[
R_Y(\theta)=
\begin{bmatrix}
\cos(\theta/2)&-\sin(\theta/2)\\
\sin(\theta/2)&\cos(\theta/2)
\end{bmatrix},
\qquad
R_Z(\theta)=
\begin{bmatrix}
e^{-i\theta/2}&0\\
0&e^{i\theta/2}
\end{bmatrix}.
\]

Single-qubit gates acting on a multi-qubit system are extended to the full Hilbert space using tensor product. For example, \(X\) acting on the first qubit and \(Z\) acting on the second qubit of a two-qubit system are represented by \(X\otimes Z\). It is common to simplify this expression to $XZ$, i.e., the tensor product symbol is omitted. 

The controlled-NOT gate (\(CX\)) is a two-qubit gate that flips the target qubit if the control qubit is in state \(|1\rangle\). With the first qubit as control and the second as target, its matrix in the ordered basis
\(\{|00\rangle,|01\rangle,|10\rangle,|11\rangle\}\) is
\[
CX=
\begin{bmatrix}
1&0&0&0\\
0&1&0&0\\
0&0&0&1\\
0&0&1&0
\end{bmatrix}.
\]
Its action on the computational basis is
\[
\begin{aligned}
CX|00\rangle&=|00\rangle, &
CX|01\rangle&=|01\rangle,\\
CX|10\rangle&=|11\rangle, &
CX|11\rangle&=|10\rangle.
\end{aligned}
\]
The \(CX\) gate can generate entanglement when applied to suitable input states and, together with single-qubit gates, forms a universal gate set for quantum computation.

\section{How the UCC Ansatz Arises from Quantum Dynamics}

A natural question in VQE is why the unitary coupled cluster (UCC) ansatz is widely used in quantum chemistry. UCC is usually introduced as a unitary extension of classical coupled cluster theory. Its form is also well suited to quantum computing, since it generates quantum states through unitary operations and allows the energy to be evaluated within a variational framework.

In classical coupled cluster theory, the wavefunction is written as an exponential of excitation operators,
\begin{equation}
|\Psi_{\mathrm{CC}}\rangle = e^{T} |\Phi_0\rangle,
\end{equation}
where \( |\Phi_0\rangle \) is a reference Slater determinant, typically obtained from a Hartree-Fock (HF) calculation, and \(T\) is the cluster operator containing particle-hole excitations. The latter is expanded in terms of excitation operators:
\begin{equation}
T = T_1 + T_2 + T_3 + \cdots,
\end{equation}
where \(T_n\) describes \(n\)-particle--\(n\)-hole excitations relative to the reference determinant.

In the coupled-cluster singles and doubles (CCSD) approximation, the expansion is truncated after the single- and double-excitation terms:
\begin{align}
T_1 &= \sum_{i,a} t_i^a\, a_a^\dagger a_i,
\label{eq:single}\\
T_2 &= \frac{1}{4}\sum_{i,j,a,b} t_{ij}^{ab}\,
a_a^\dagger a_b^\dagger a_j a_i.
\end{align}
Here, \(i,j\) label occupied orbitals in the reference determinant, while \(a,b\) label unoccupied (virtual) orbitals. The coefficients \(t_i^a\) and \(t_{ij}^{ab}\), known as cluster amplitudes, determine the contributions of the corresponding excitations to the wavefunction.

The exponential form accounts for electron correlation and gives coupled cluster theory its size-extensive character. However, $T$ is generally not anti-Hermitian, making $e^T$ non-unitary. Although the resulting wavefunction can be normalized, the operator $e^T$ does not describe unitary quantum evolution. Furthermore, the standard coupled cluster energy is not obtained from a variational expectation value, so it is not guaranteed to lie above the exact ground-state energy.

In quantum circuits, state evolution is described by unitary operators. This motivates replacing the classical coupled cluster operator $T$ with the anti-Hermitian operator $T-T^\dagger$, giving the UCC ansatz
\begin{equation}
|\Psi_{\mathrm{UCC}}\rangle = e^{T - T^\dagger} |\Phi_0\rangle,
\end{equation}
where $(T-T^\dagger)^\dagger=-(T-T^\dagger)$. The resulting exponential is unitary and can therefore be used to prepare quantum states on a quantum computer. UCC retains the exponential form of classical coupled cluster theory while ensuring that the state transformation is unitary.

The difference between the two approaches also affects how their parameters are determined. In classical coupled cluster theory, the amplitudes are usually obtained by solving projective equations. In VQE, the parameters of the UCC ansatz are instead optimized to minimize the energy expectation value. The ansatz thus defines a family of states that can be prepared by quantum circuits, and the quality of the energy estimate depends on how well this family represents the ground state.

The operator $T-T^\dagger$ is built from fermionic excitation and de-excitation operators. Each term connects Slater determinants that differ by one or more orbital occupations. After applying a fermion-to-qubit mapping such as the Jordan--Wigner transformation, these operators can be written as sums of Pauli operators. Their exponentials can then be decomposed into quantum gates. This provides a direct route from the fermionic description to a quantum circuit.

A practical difficulty arises because the individual excitation and de-excitation terms generally do not commute. Consequently, the exponential of their sum cannot, in general, be expressed exactly as a product of the corresponding exponentials:
\begin{equation}
    e^{A+B} \not= e^Ae^B \qquad [A, B] \not= 0
\end{equation}
where $[A,B]=AB-BA$ denotes the commutator. For an exposition on the issue of commutativity in coupled cluster theory see Appendices~\ref{app:A} and ~\ref{app:i}.

A common solution is to use a Trotter decomposition, which approximates the unitary by an ordered product of simpler operations. For example,
\begin{equation}
e^{\sum_k \tau_k}
\approx
\prod_k e^{\tau_k},
\end{equation}
where the approximation becomes exact in the limit of an increasing number of Trotter steps under the usual conditions. Here, each $\tau_k$ is an anti-Hermitian operator associated with an excitation or de-excitation term. 

Due to non-commutativity, the order of the $\tau_k$ operators matters when approximating the UCCSD unitary using a Trotter decomposition. Different orderings can produce different circuits and, at a finite Trotter depth, different quantum states. In the limit of infinitely many Trotter steps, these approximations converge to the same unitary, provided they represent the same generator. Thus, although the exact UCCSD ansatz is well defined, its circuit implementation is not unique. The choice of operator ordering, decomposition, and number of Trotter steps can affect circuit depth, approximation error, and VQE optimization performance. These considerations are particularly important for UCCSD, which includes both single and double excitations and must be implemented with a circuit that remains practical for quantum hardware.

The UCC ansatz used in VQE can be viewed as a compromise between three requirements: (i) retaining the excitation structure of many-body theory, (ii) ensuring unitary evolution, and (iii) allowing implementation as a quantum circuit. Its form builds on classical coupled cluster theory while adapting the ansatz to the constraints of quantum computation.

\section{\label{sec:motivation}Motivation for the Jordan-Wigner mapping}

The Jordan–Wigner (JW) transformation \cite{JordanWigner1928} maps fermionic operators to qubit operators. This allows electronic-structure Hamiltonians to be simulated on quantum computers. Other mappings, including Bravyi–Kitaev \cite{bravyi}, parity \cite{bravyi}, and superfast encodings \cite{superfast, ultrafast}, offer different trade-offs in level of intuitiveness, operator complexity and circuit cost. We use the JW transformation here because its construction is straightforward and provides a clear starting point for understanding fermion-to-qubit mappings.

The JW transformation preserves the anticommutation relations of fermionic creation and annihilation operators when they are represented by qubit operators. Operators acting on different qubits commute because they act on separate subsystems, whereas fermionic creation and annihilation operators obey anticommutation relations. The JW mapping accounts for this difference by introducing sequences of Pauli-$Z$ operators. These operator sequences encode the parity of the occupations of preceding orbitals, which ensures that the mapped operators satisfy the correct fermionic anticommutation relations.

The mapping also establishes a direct correspondence between spin-orbitals and qubits, with each qubit representing the occupation of one spin-orbital. This makes it relatively easy to follow how a molecular Hamiltonian or an excitation operator is translated into qubit form. In the context of UCCSD, the same procedure can be used to express fermionic excitation generators in terms of Pauli operators and then construct the corresponding quantum circuits. This provides a useful basis for understanding other mappings, such as Bravyi–Kitaev, which encode occupation information differently and can offer advantages in operator locality and circuit implementation.


For a spin-orbital indexed by $p$, the JW transformation maps the fermionic annihilation and creation operators to qubit operators as
\begin{equation}\label{eq:jw-map}
\begin{split}
a_p
&=
\frac{1}{2}
\left(
X_p + iY_p
\right)
\prod_{j=0}^{p-1} Z_j,
\\
a_p^\dagger
&=
\frac{1}{2}
\left(
X_p - iY_p
\right)
\prod_{j=0}^{p-1} Z_j,
\end{split}
\end{equation}
where $X_p$, $Y_p$, and $Z_p$ are Pauli operators acting on qubit $p$. For $p=0$, the product is understood to be the identity operator.

The need for this mapping comes from the different algebraic properties of qubit and fermionic operators. Qubits represent distinguishable two-level systems, while electrons are fermions and obey the fermionic anticommutation relations.

Operators acting on different qubits commute. For example, applying $X_0$ and $X_1$ in either order gives the same result:
\begin{equation}
[X_0,X_1]=0
\quad\implies\quad
X_0X_1=X_1X_0,
\end{equation}

Fermionic creation operators, in contrast, anticommute, so exchanging the order of two such operators changes the sign:
\begin{equation}
\{a_0^\dagger,a_1^\dagger\}=0
\quad\implies\quad
a_0^\dagger a_1^\dagger
=-a_1^\dagger a_0^\dagger,
\end{equation}
where ${A,B}=AB+BA$ denotes the anticommutator. This relation reflects the antisymmetry of fermionic states and is closely related to the Pauli exclusion principle. The JW transformation preserves this property by including a sequence of Pauli-$Z$ operators in each mapped creation and annihilation operator.

To see why this is necessary, consider representing each spin-orbital with a qubit, where $|0\rangle$ denotes an empty orbital and $|1\rangle$ an occupied orbital. A mapping based only on local qubit operations would not reproduce the anticommutation relations between different orbitals. The JW transformation addresses this by writing the creation operator as
\begin{equation}
    a_p^\dagger = \underbrace{\frac{1}{2}(X_p - iY_p)}_{\text{local state change}} \otimes \underbrace{\prod_{j=0}^{p-1}Z_j}_{\text{parity / phase correction}}
\end{equation}
Here, $(X_p-iY_p)/2$ changes the occupation of qubit $p$ from $|0\rangle$ to $|1\rangle$, while the $Z$ operators account for the parity of the occupations of all preceding orbitals. This parity factor supplies the sign needed to reproduce fermionic anticommutation relations.

In particular, $(X-iY)/2$ acts as a raising operator under this occupation convention, while $(X+iY)/2$ acts as a lowering operator. They therefore represent the local occupation changes associated with creation and annihilation, respectively. The parity string completes the mapping by ensuring that these local operations obey the correct fermionic algebra.


The raising operator flips the state of the \(p\)-th qubit from \(|0\rangle\) (an empty spin orbital) to \(|1\rangle\) (an occupied spin orbital). Under this occupation convention, the raising operator is defined as
\[
\sigma^+ = \frac{X-iY}{2}.
\]
Its matrix representation is
\[
\begin{aligned}
\sigma^+
&=\frac{1}{2}\left[
\begin{bmatrix}0&1\\1&0\end{bmatrix}
-i\begin{bmatrix}0&-i\\i&0\end{bmatrix}
\right]
=\begin{bmatrix}0&0\\1&0\end{bmatrix}  =|1\rangle\langle 0|
\end{aligned}
\]

The lowering operator changes the state of the \(p\)-th qubit from \(|1\rangle\) (an occupied spin orbital) to \(|0\rangle\) (an empty spin orbital). Under this occupation convention, it is defined as
\[
\sigma^-=\frac{X+iY}{2}.
\]
Its matrix representation is
\[
\begin{aligned}
\sigma^-
&=\frac{1}{2}\left[
\begin{bmatrix}0&1\\1&0\end{bmatrix}
+i\begin{bmatrix}0&-i\\i&0\end{bmatrix}
\right]
=\begin{bmatrix}0&1\\0&0\end{bmatrix}=|0\rangle\langle 1|.
\end{aligned}
\]

Both the raising and lowering operators update the occupation of a spin orbital correctly through the combination of the Pauli \(X\) and \(Y\) operators. The relative phase between their contributions ensures that the operator acts only on the appropriate occupation state.

For the raising operator \(\sigma^+\) the action on the empty state \(|0\rangle\) is
\[
\sigma^+|0\rangle
=\frac{1}{2}\left(X|0\rangle-iY|0\rangle\right)
=\frac{1}{2}\left(|1\rangle+|1\rangle\right)
=|1\rangle.
\]
Here, \(X|0\rangle=|1\rangle\) and \(iY|0\rangle=-|1\rangle\), so the two contributions add constructively in \(X-iY\). In contrast, when the raising operator acts on the occupied state \(|1\rangle\),
\[
\sigma^+|1\rangle
=\frac{1}{2}\left(X|1\rangle-iY|1\rangle\right)
=\frac{1}{2}\left(|0\rangle-|0\rangle\right)
=0,
\]
because \(X|1\rangle=|0\rangle\) and \(iY|1\rangle=|0\rangle\). The cancellation ensures that the creation operator vanishes when the spin orbital is already occupied, consistent with the Pauli exclusion principle.

Similarly, the lowering operator \(\sigma^-\) annihilates the empty state:
\[
\sigma^-|0\rangle
=\frac{1}{2}\left(X|0\rangle+iY|0\rangle\right)
=\frac{1}{2}\left(|1\rangle-|1\rangle\right)
=0.
\]
The two contributions cancel, so the operator cannot remove a particle from an empty spin orbital. When it acts on the occupied state, however,
\[
\sigma^-|1\rangle
=\frac{1}{2}\left(X|1\rangle+iY|1\rangle\right)
=\frac{1}{2}\left(|0\rangle+|0\rangle\right)
=|0\rangle.
\]
In this case, the contributions add constructively, leaving the spin orbital empty after the particle is removed.

The nonlocal part of the JW transformation, commonly referred to as the \(Z\)-string, ensures that the mapped operators satisfy the fermionic anticommutation relations by encoding the parity of the occupations of all preceding spin orbitals. The Pauli-\(Z\) operator acts on the computational basis states as
\[
\begin{aligned}
Z|0\rangle &= +|0\rangle, \qquad &&\text{(empty spin orbital)},\\
Z|1\rangle &= -|1\rangle, &&\text{(occupied spin orbital)}.
\end{aligned}
\]
Thus, applying \(Z\) to an occupied orbital contributes a factor of \(-1\), whereas applying it to an empty orbital leaves the state unchanged.

For the \(p\)-th spin orbital, the JW parity string is
\[
P_p=\prod_{j=0}^{p-1}Z_j.
\]
If \(n_j\in\{0,1\}\) denotes the occupation number of orbital \(j\), then its action on a computational basis state contributes the phase
\[
P_p|n_0n_1\cdots n_{N-1}\rangle
=(-1)^{\sum_{j=0}^{p-1}n_j}
|n_0n_1\cdots n_{N-1}\rangle.
\]
The resulting phase is \(+1\) when the number of occupied preceding orbitals is even and \(-1\) when it is odd. The \(Z\)-string therefore encodes the fermionic parity associated with the orbitals preceding \(p\).


The role of the \(Z\)-string can be illustrated using two spin orbitals:
\[
a_0^\dagger=\frac{X_0-iY_0}{2},\qquad
a_1^\dagger=\frac{X_1-iY_1}{2}Z_0.
\]
Starting from the vacuum state \(|00\rangle\), creating a particle in orbital \(0\) followed by one in orbital \(1\) is represented by \(a_1^\dagger a_0^\dagger\). The process involves two operations. First, a particle is created in orbital \(0\):
\[
a_0^\dagger|00\rangle=|10\rangle.
\]
Then, a particle is created in orbital \(1\):
\[
a_1^\dagger|10\rangle=-|11\rangle,
\]
because \(Z_0\) contributes a factor of \(-1\) when orbital \(0\) is occupied. In the reverse order,
\[
a_0^\dagger a_1^\dagger|00\rangle=|11\rangle,
\]
since orbital \(0\) is initially empty when \(a_1^\dagger\) acts. The opposite signs demonstrate that
\[
\{a_0^\dagger,a_1^\dagger\}
=a_0^\dagger a_1^\dagger+a_1^\dagger a_0^\dagger=0,
\]
as required by fermionic anticommutation. More generally, the \(Z\)-string encodes the parity of preceding orbital occupations, allowing the JW transformation to reproduce fermionic statistics using qubit operators.

\section{Example: excitations in the hydrogen molecule}
\label{sec:example}

We use Qiskit~\cite{qiskit2024} to illustrate the construction of excitation operators for the hydrogen molecule (\(H_2\)). In the minimal STO-3G basis, \(H_2\) has four spin orbitals. Qiskit Nature adopts a blocked spin-orbital ordering, where all \(\alpha\)-spin orbitals are listed before the \(\beta\)-spin orbitals: \[ (\alpha_0,\alpha_1,\beta_0,\beta_1). \] Here, subscripts \(0\) and \(1\) label the bonding and antibonding spatial orbitals, respectively. The corresponding spin-orbital indices are therefore \begin{itemize} \item 0, 1: \(\alpha\)-spin orbitals \begin{itemize} \item 0: bonding (occupied) \item 1: antibonding (virtual) \end{itemize} \item 2, 3: \(\beta\)-spin orbitals \begin{itemize} \item 2: bonding (occupied) \item 3: antibonding (virtual) \end{itemize} \end{itemize} This convention differs from the interleaved ordering used in some quantum-computing frameworks, where the two spin orbitals associated with each spatial orbital are placed consecutively.

\subsection{Single excitations}

A single excitation transfers an electron from an occupied orbital \(i\) to an unoccupied (virtual) orbital \(a\). In the Hartree-Fock reference state for \(H_2\), orbitals \(0\) and \(2\) are occupied, while orbitals \(1\) and \(3\) are virtual. 

Two constraints determine the allowed excitations: 

\paragraph{Spin conservation} In standard non-relativistic quantum chemistry, excitation operators preserve spin. An \(\alpha\)-spin electron must remain \(\alpha\), and a \(\beta\)-spin electron must remain \(\beta\). Therefore, transitions such as \(0\rightarrow3\) and \(2\rightarrow1\) are forbidden because they would flip the spin. 

\paragraph{Pauli exclusion principle} An excitation must move an electron from an occupied orbital to an unoccupied one. Thus, transitions such as \(0\rightarrow2\) are not allowed because orbital \(2\) is already occupied, and no excitation can originate from orbitals \(1\) or \(3\) because they are empty. Combining these two rules leaves only two spin-conserving single excitations: 
\begin{align} 
a_1^\dagger a_0, \qquad a_3^\dagger a_2. 
\end{align} 

The operator \(a_1^\dagger a_0\) excites an \(\alpha\)-spin electron from the bonding orbital to the antibonding orbital, while \(a_3^\dagger a_2\) performs the corresponding excitation for a \(\beta\)-spin electron. These are the only occupied-to-virtual single excitations available for this reference configuration.

\subsection{Double excitation} 

A double excitation simultaneously promotes two electrons from occupied orbitals to virtual orbitals. As for single excitations, a double excitation must satisfy both spin conservation and the Pauli exclusion principle. In the Hartree-Fock reference state of \(H_2\), orbitals \(0\) and \(2\) are occupied, while orbitals \(1\) and \(3\) are virtual. Since spin must be conserved, the \(\alpha\)-spin electron in orbital \(0\) can only be excited to orbital \(1\), and the \(\beta\)-spin electron in orbital \(2\) can only be excited to orbital \(3\).

There is therefore only one possible spin-conserving double excitation: 
\begin{align} 
a_1^\dagger a_3^\dagger a_0 a_2. 
\end{align}
This operator removes the \(\alpha\)-spin electron from orbital \(0\) and the \(\beta\)-spin electron from orbital \(2\), and places them in the antibonding orbitals \(1\) and \(3\), respectively. No other double excitations satisfy both spin conservation and the requirement that electrons be excited from occupied to virtual orbitals. Any alternative choice would either place an electron into an already occupied orbital, originate from a virtual orbital, or require a spin flip. Consequently, \(a_1^\dagger a_3^\dagger a_0 a_2\) is the only double-excitation operator for \(H_2\) in the STO-3G basis.

 The ordering of the creation operators follows the convention used in Qiskit Nature. Other texts may write the equivalent excitation operator as \( a_3^\dagger a_1^\dagger a_2 a_0 \), which differs only by operator ordering and an overall sign arising from the fermionic anticommutation relations.

\subsection{Example mapping of single and double excitations}

Under the JW mapping, for a spin-orbital index $p$, the fermionic operators map to qubit operators as given in Eq.~(\ref{eq:jw-map}). We illustrate the procedure for both single and double excitations in the \(H_2\) molecule.

\subsubsection{Single excitation}

Consider the excitation of an \(\alpha\)-spin electron from the occupied bonding orbital \(0\) to the virtual antibonding orbital \(1\). The corresponding fermionic operator is $a_1^\dagger a_0$.

The JW mapping proceeds as follows. 
\begin{enumerate} \item Map the annihilation operator \(a_0\). 

Since orbital \(0\) has no preceding orbitals, no \(Z\)-string is required: 
\begin{equation} 
a_0=\frac{1}{2}\left(X_0+iY_0\right). 
\end{equation} 

\item Map the creation operator \(a_1^\dagger\). 

For orbital \(1\), the string contains a \(Z\) operator acting on qubit \(0\): 
\begin{equation} 
a_1^\dagger=\frac{1}{2}\left(X_1-iY_1\right)Z_0. \end{equation} 

\item Form the product. 

Combining the two mapped operators gives 
\begin{equation} 
a_1^\dagger a_0 =\frac{1}{4} \left(X_1-iY_1\right) Z_0 \left(X_0+iY_0\right). 
\end{equation} 

\item Simplify the expression. 

Using the Pauli identities 
\begin{equation} 
Z_0X_0=iY_0, \qquad Z_0Y_0=-iX_0, 
\end{equation} 
we find 
\begin{align} 
Z_0(X_0+iY_0) &=iY_0+i(-iX_0)\\ &=X_0+iY_0. 
\end{align} 

Therefore, 
\begin{equation} 
a_1^\dagger a_0 = \frac{1}{4} \left(X_1-iY_1\right) \left(X_0+iY_0\right). 
\end{equation} 

\item Expand the product. 
\begin{align} 
a_1^\dagger a_0 = \frac{1}{4} \Big( X_1X_0 +iX_1Y_0 -iY_1X_0 +Y_1Y_0 \Big). 
\end{align} 

\end{enumerate} 

The single-excitation operator is therefore mapped to a sum of four Pauli operators: 
\begin{equation}
    a_1^\dagger a_0 = \frac{1}{4} (X_1 X_0 + Y_1 Y_0) + \frac{i}{4} (X_1 Y_0 - Y_1 X_0)
\end{equation}

\subsubsection{Double excitation}

The only double excitation in the STO-3G description of \(H_2\) simultaneously promotes the \(\alpha\)- and \(\beta\)-spin electrons from the bonding orbitals to the antibonding orbitals: $a_1^\dagger a_3^\dagger a_0 a_2$. Applying the JW transformation gives 
\begin{align} 
a_0 &= \frac{1}{2} (X_0+iY_0), \\ a_1^\dagger &= \frac12 (X_1-iY_1)Z_0, \\ a_2 &= \frac12 (X_2+iY_2)Z_1Z_0, \\ a_3^\dagger &= \frac12 (X_3-iY_3)Z_2Z_1Z_0. 
\end{align} 

It is convenient to first evaluate the \(\beta\)-spin excitation operator: \begin{align} a_3^\dagger a_2 &= \frac{1}{4} (X_3-iY_3) Z_2Z_1Z_0 (X_2+iY_2) Z_1Z_0 \\ &= \frac{1}{4} (X_3-iY_3) Z_2 (X_2+iY_2), \end{align} since \(Z_1^2=Z_0^2=I\). Using \[ Z_2X_2=iY_2, \qquad Z_2Y_2=-iX_2, \] we obtain \begin{equation} a_3^\dagger a_2 = \frac{1}{4} \left( X_3X_2+Y_3Y_2 \right) + \frac{i}{4} \left( X_3Y_2-Y_3X_2 \right). \end{equation} The corresponding \(\alpha\)-spin excitation is \begin{equation} a_1^\dagger a_0 = \frac{1}{4} \left( X_1X_0+Y_1Y_0 \right) + \frac{i}{4} \left( X_1Y_0-Y_1X_0 \right). \end{equation} The full double-excitation operator is therefore 
\begin{equation}\label{16-terms}
a_1^\dagger a_3^\dagger a_0 a_2 = (a_1^\dagger a_0)(a_3^\dagger a_2), 
\end{equation} 
which expands into sixteen four-qubit Pauli operators.


\section{Quantum circuit implementation for single excitations \label{sec:circuit}}

To implement a single excitation in a quantum circuit, the anti-Hermitian excitation operator is exponentiated to obtain a unitary transformation. For the excitation \(a_1^\dagger a_0\), the UCCSD ansatz uses 
\begin{equation} 
U(\theta) = e^{T-T^\dagger} = e^{\theta (a_1^\dagger a_0-a_0^\dagger a_1)}, 
\end{equation} 
where \(\theta=t_0^1\) is the variational parameter associated with this excitation [see Eq.~(\ref{eq:single})]. 

Using the JW transformation, the excitation and de-excitation operators are 
\begin{align} 
a_1^\dagger a_0 &= \frac14 (X_1X_0+Y_1Y_0) +\frac{i}{4}(X_1Y_0-Y_1X_0),\\ a_0^\dagger a_1 &= \frac14 (X_1X_0+Y_1Y_0) -\frac{i}{4}(X_1Y_0-Y_1X_0). 
\end{align} 

Taking the difference gives 
\begin{equation} 
a_1^\dagger a_0-a_0^\dagger a_1 = \frac{i}{2} \left( X_1Y_0-Y_1X_0 \right). 
\end{equation} 

The unitary operator therefore becomes 
\begin{equation} 
U(\theta) = \exp\!\left[ i\frac{\theta}{2} \left( X_1Y_0-Y_1X_0 \right) \right]. 
\end{equation} 

The generator consists of two Pauli operators, \(X_1Y_0\) and \(Y_1X_0\), which can be implemented using standard Pauli-rotation circuits.

A useful feature of the JW mapping is that the generator of the single excitation becomes a sum of two commuting Pauli operators: 
\begin{equation} 
a_1^\dagger a_0-a_0^\dagger a_1 = \frac{i}{2} \left( X_1Y_0-Y_1X_0 \right). 
\end{equation} 

The two Pauli operators commute, 
\begin{equation} 
[X_1Y_0,\;Y_1X_0]=0, 
\end{equation} 
which can be verified directly using the Pauli algebra. As a result, the exponential may be factorized exactly: 
\begin{align} 
U(\theta) &= \exp\!\left[ i\frac{\theta}{2} \left( X_1Y_0-Y_1X_0 \right) \right] \\ &= e^{\,i\frac{\theta}{2}X_1Y_0} e^{-i\frac{\theta}{2}Y_1X_0}. 
\end{align} 

No Trotter approximation is required because the two Pauli operators commute exactly. Each exponential represents a Pauli rotation and can be implemented using standard basis-change gates, a controlled \(R_z\) rotation, and the inverse basis transformation. The angle supplied to the hardware rotation gate is determined by matching the desired unitary to the convention \begin{equation} R_z(\lambda)=e^{-i\lambda Z/2}. \end{equation} Consequently, the coefficient appearing in the UCC generator must be translated into the corresponding circuit rotation angle during the circuit synthesis step.

To implement an operator of the form 
\begin{equation} 
e^{-i\phi(P_1 \otimes P_0)}, 
\end{equation} 
where \(P_k \in \{X,Y,Z\}\), a standard circuit construction is used: 
\begin{enumerate} 
\item Basis change. 

Rotate each qubit so that the corresponding Pauli operator is mapped to \(Z\): 
\begin{itemize} 
\item \(X \rightarrow Z\): apply a Hadamard gate \(H\), \item \(Y \rightarrow Z\): apply \(S^\dagger H\) (or equivalently \(R_x(\pi/2)\) up to a phase), 
\item \(Z \rightarrow Z\): no gate is required. \end{itemize} 

\item Parity computation. 

Apply a sequence of CX gates to encode the parity of the Pauli operator onto a target qubit. 

\item Rotation.

Apply 
\begin{equation} 
R_z(2\phi) 
\end{equation} 
to the target qubit. Since $R_z(\lambda)=e^{-i\lambda Z/2}$, the choice \(\lambda=2\phi\) implements the desired phase factor. 

\item Uncomputation. 

Reverse the CX sequence and undo the basis-change gates to return to the original basis. 
\end{enumerate}

We now construct the circuits for the two commuting Pauli rotations, 
\begin{equation} 
e^{i\frac{\theta}{2}X_1Y_0}, \qquad e^{-i\frac{\theta}{2}Y_1X_0}. 
\end{equation} 

\subsection{Term \(e^{i\frac{\theta}{2}X_1Y_0}\)} 

To map \(X_1Y_0\) onto a \(Z\)-type parity string, qubit 1 is rotated from the \(X\)-basis to the \(Z\)-basis and qubit 0 is rotated from the \(Y\)-basis to the \(Z\)-basis. Since $R_z(\lambda)=e^{-i\lambda Z/2}$, the positive sign in the target exponent requires a rotation angle of \(-\theta\). \begin{enumerate} \item Basis change: \begin{itemize} \item Qubit 0 (\(Y\)): apply \(R_x(\pi/2)\). \item Qubit 1 (\(X\)): apply \(H\). \end{itemize} \item Parity computation: Apply \(\mathrm{CX}(0 \rightarrow 1)\). \item Rotation: Apply \(R_z(-\theta)\) to qubit 1. \item Uncompute: \begin{itemize} \item Apply \(\mathrm{CX}(0 \rightarrow 1)\). \item Apply \(H\) to qubit 1. \item Apply \(R_x(-\pi/2)\) to qubit 0. \end{itemize} \end{enumerate} 

\subsection{Term \(e^{-i\frac{\theta}{2}Y_1X_0}\)} For \(Y_1X_0\), qubit 1 is rotated from the \(Y\)-basis to the \(Z\)-basis and qubit 0 from the \(X\)-basis to the \(Z\)-basis. The sign in the exponent already matches the convention of the \(R_z\) gate, so the rotation angle is \(+\theta\). \begin{enumerate} \item Basis change: \begin{itemize} \item Qubit 0 (\(X\)): apply \(H\). \item Qubit 1 (\(Y\)): apply \(R_x(\pi/2)\). \end{itemize} \item Parity computation: Apply \(\mathrm{CX}(0 \rightarrow 1)\). \item Rotation: Apply \(R_z(\theta)\) to qubit 1. \item Uncompute: \begin{itemize} \item Apply \(\mathrm{CX}(0 \rightarrow 1)\). \item Apply \(R_x(-\pi/2)\) to qubit 1. \item Apply \(H\) to qubit 0. \end{itemize} \end{enumerate}

\subsection{Handling nonlocal excitations (Z-strings)}

The \(H_2\) example considered in this chapter contains only nearest-neighbor single excitations and therefore does not require Jordan--Wigner \(Z\)-strings. Nevertheless, longer \(Z\)-strings appear in larger molecules whenever an excitation connects non-adjacent spin orbitals. For example, a nonlocal excitation may give rise to a Pauli string of the form \[ X_2 Z_1 Y_0. \] The intermediate \(Z\)-operators encode fermionic parity and must be included in the circuit implementation. To implement \[ e^{-i\phi\, X_2 Z_1 Y_0}, \] the standard Pauli-string construction is extended as follows: 
\begin{enumerate} 
\item Basis change. 

Rotate only the endpoint qubits into the \(Z\)-basis: \begin{itemize} 
\item Qubit \(0\) (\(Y\)): apply \(R_x(\pi/2)\). 
\item Qubit \(2\) (\(X\)): apply \(H\). 
\item Qubit \(1\) (\(Z\)): no basis change is required. \end{itemize} 

\item Parity computation. 

Compute the parity of the entire Pauli string using a CX ladder: \[ \mathrm{CX}(0\rightarrow1), \qquad \mathrm{CX}(1\rightarrow2). \] 

\item Rotation. 

Apply \(R_z(2\phi)\) to the final target qubit (\(2\)). 

\item Uncomputation. 

Reverse the CX sequence and undo the basis-change gates. 
\end{enumerate} 

In general, a JW $Z$-string spanning \(n\) qubits requires a chain of \(n-1\) CX gates to compute the parity, followed by the corresponding \(R_z\) rotation and the inverse CX chain.

\subsection{Remarks on the implementation}

\subsubsection{Direction of CX}

To implement the Pauli rotations \[ U_1=e^{-i\frac{\theta}{2}Y_1X_0} \qquad\text{or}\qquad U_2=e^{i\frac{\theta}{2}X_1Y_0}, \] the first step is to rotate the Pauli operators into the \(Z\)-basis. We illustrate the procedure using $U_2$. \begin{itemize} \item Qubit \(q_1\): \(X \mapsto Z\) via a Hadamard gate \(H\). \item Qubit \(q_0\): \(Y \mapsto Z\) via \(R_x(\pi/2)\). \end{itemize} 

After the basis transformation, the problem reduces to implementing \[ e^{i\frac{\theta}{2}Z_1Z_0}. \] The parity \(Z_1Z_0\) can be computed using either qubit as the target of the CX chain. Consequently, the direction of the CX gate is not unique, and different software frameworks may choose different conventions. Since the target exponent is positive and Qiskit's native rotation gate is defined as \[ R_z(\lambda)=e^{-i\lambda Z/2}, \] the required rotation angle is \(-\theta\). 

\subparagraph{Option 1 (target qubit \(q_1\))} \begin{equation*} \Qcircuit @C=1em @R=1em { \lstick{q_0 (Y)} & \gate{R_x(\frac{\pi}{2})} & \ctrl{1} & \qw & \ctrl{1} & \gate{R_x(-\frac{\pi}{2})} & \qw \\ \lstick{q_1 (X)} & \gate{H} & \targ & \gate{R_z(-\theta)} & \targ & \gate{H} & \qw } \end{equation*} 

\subparagraph{Option 2 (target qubit \(q_0\), used by Qiskit)} \begin{equation*} \Qcircuit @C=1em @R=1em { \lstick{q_0 (Y)} & \gate{R_x(\frac{\pi}{2})} & \targ & \gate{R_z(-\theta)} & \targ & \gate{R_x(-\frac{\pi}{2})} & \qw \\ \lstick{q_1 (X)} & \gate{H} & \ctrl{-1} & \qw & \ctrl{-1} & \gate{H} & \qw } \end{equation*} 

Both circuits implement the same unitary operator, \[ e^{i\frac{\theta}{2}X_1Y_0}. \] The parameter \(\theta\) is the variational parameter optimized during the VQE procedure. In the UCCSD ansatz, it corresponds to the excitation amplitude \(t_a^i\) appearing in the cluster operator \(T\) [Eq.~(\ref{eq:single})].

\subsubsection{Comparison: $R_x(\pi/2)$ vs $\sqrt{X}$    \label{subsec:sx}}

As shown in Fig.~\ref{fig:h2-2nd-dec}, Qiskit Nature uses $\sqrt{X}$ instead of the $R_X(\pi/2)$ gate, and $\sqrt{X}^\dagger$ instead of $R_x(-\pi/2)$. Strictly speaking, \(R_x(\pi/2)\) and \(\sqrt{X}\) are not identical matrices. They differ by a global phase. The rotation operator \(R_x(\pi/2)\) is defined as \begin{equation} R_x(\pi/2) = e^{-i\frac{\pi}{4}X} = \frac{1}{\sqrt{2}} \begin{pmatrix} 1 & -i \\ -i & 1 \end{pmatrix}. \end{equation} Squaring this operator gives \begin{equation} \bigl(R_x(\pi/2)\bigr)^2 = -iX. \end{equation} 

The \(\sqrt{X}\) gate is defined as the principal square root of the Pauli-\(X\) operator: \begin{equation} \sqrt{X} = \frac{1}{2} \begin{pmatrix} 1+i & 1-i \\ 1-i & 1+i \end{pmatrix}. \end{equation} By construction, \begin{equation} \bigl(\sqrt{X}\bigr)^2 = X. \end{equation} The two gates are related by \begin{equation} R_x(\pi/2) = e^{-i\pi/4}\sqrt{X}. \end{equation} Therefore, they produce the same rotation on the Bloch sphere and differ only by a global phase of \(-\pi/4\). For single-qubit operations, this distinction is physically irrelevant because global phases do not affect measurement probabilities. Since measurement outcomes depend only on the squared amplitudes of the quantum state, the phase factor cancels. Consequently, \(R_x(\pi/2)\) and \(\sqrt{X}\) can be used interchangeably when performing basis changes, such as mapping the \(Y\)-basis onto the \(Z\)-basis. 

The situation is different for controlled operations. Although the global phase of an isolated gate is unobservable, promoting that gate to a controlled operation generally converts the global phase into a relative phase between different components of the joint quantum state. As a result, controlled-\(R_x(\pi/2)\) and controlled-\(\sqrt{X}\) are not equivalent and can produce different observable outcomes. In the JW circuits considered in this work, \(R_x(\pi/2)\) is used to describe the basis transformation at the algorithmic level. On superconducting quantum hardware, however, this operation is often implemented using the native \(\sqrt{X}\) (SX) gate together with phase corrections supplied by the compiler.

\section{Quantum circuit implementation for double excitation}

The only double excitation in the STO-3G description of \(H_2\) is \begin{equation} T_2=a_1^\dagger a_3^\dagger a_0 a_2. \end{equation} The corresponding UCCSD unitary is \begin{equation} U(\theta) = e^{\theta(T_2-T_2^\dagger)}, \end{equation} where \(\theta=t_{02}^{13}\) is the variational parameter associated with the double excitation, and 
\begin{equation}
    T^\dagger_2 = (a_1^\dagger a_3^\dagger a_0 a_2)^\dagger = a_2^\dagger a_0^\dagger a_3a_1.
\end{equation}

Applying the JW transformation as in Eq.~(\ref{16-terms}) and forming the anti-Hermitian combination gives 8 terms
\begin{align} 
T_2-T_2^\dagger = \frac{i}{8}\Big( & Y_0X_1X_2X_3 - X_0Y_1X_2X_3 \nonumber\\
&+ X_0X_1Y_2X_3 - X_0X_1X_2Y_3 \nonumber\\ 
& - X_0Y_1Y_2Y_3 + Y_0X_1Y_2Y_3 \nonumber\\
&- Y_0Y_1X_2Y_3 + Y_0Y_1Y_2X_3 \Big). 
\end{align} 



\section{Analysis of the UCCSD circuit for $H_2$ in Qiskit Nature}

\begin{figure*}
    \includegraphics[width=0.95\linewidth]{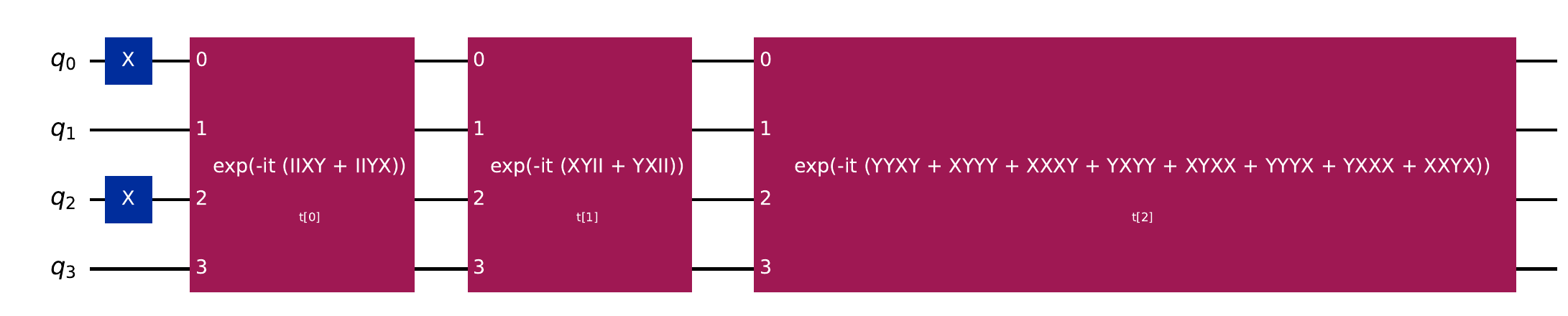}
    \caption{Qiskit circuit for the $H_2$ molecule decomposed to show the unitary evolution operators. It comprises the Hartree-Fock state and the single and double excitation operators. Parameter $t[0]$ refers to the coefficient of $a^\dagger_1 a_0$, parameter $t[1]$ refers to the coefficient of $a^\dagger_3 a_2$, while parameter $t[2]$ is the coefficient of the double-excitation term $a^\dagger_1 a^\dagger_3 a_0 a_2$.}
    \label{fig:h2-1st-dec}
\end{figure*}

Fig.~\ref{fig:h2-1st-dec} illustrates the quantum circuit for $H_2$ decomposed at the level of unitary evolution operators. The reference Hartree-Fock state contains two electrons occupying the lowest-energy bonding orbital with opposite spins: 
\begin{equation}
    |\Phi_0\rangle =a_0^\dagger a_2^\dagger |00\rangle.
\end{equation}
The occupation vector is therefore
\begin{equation}
    |1010\rangle
\end{equation}
Qiskit, however, displays computational basis states in reversed bit order. Consequently, the same state appears as $|q_3 q_2 q_1 q_0\rangle = |0101\rangle$ when read directly from circuit output.

The operators shown inside the exponential blocks correspond to the anti-Hermitian generators obtained from the JW mapping in Section~\ref{sec:example}. Parameter \(t[0]\) is associated with the \(\alpha\)-spin single excitation \(a_1^\dagger a_0\), while \(t[1]\) corresponds to the \(\beta\)-spin single excitation \(a_3^\dagger a_2\). The parameter \(t[2]\) controls the double excitation \(a_1^\dagger a_3^\dagger a_0 a_2\). At this level of decomposition, the circuit retains a direct correspondence with the underlying fermionic excitation operators, allowing the spin and particle-number symmetries to be verified before further circuit synthesis.

\begin{figure*}
    \includegraphics[width=0.95\linewidth]{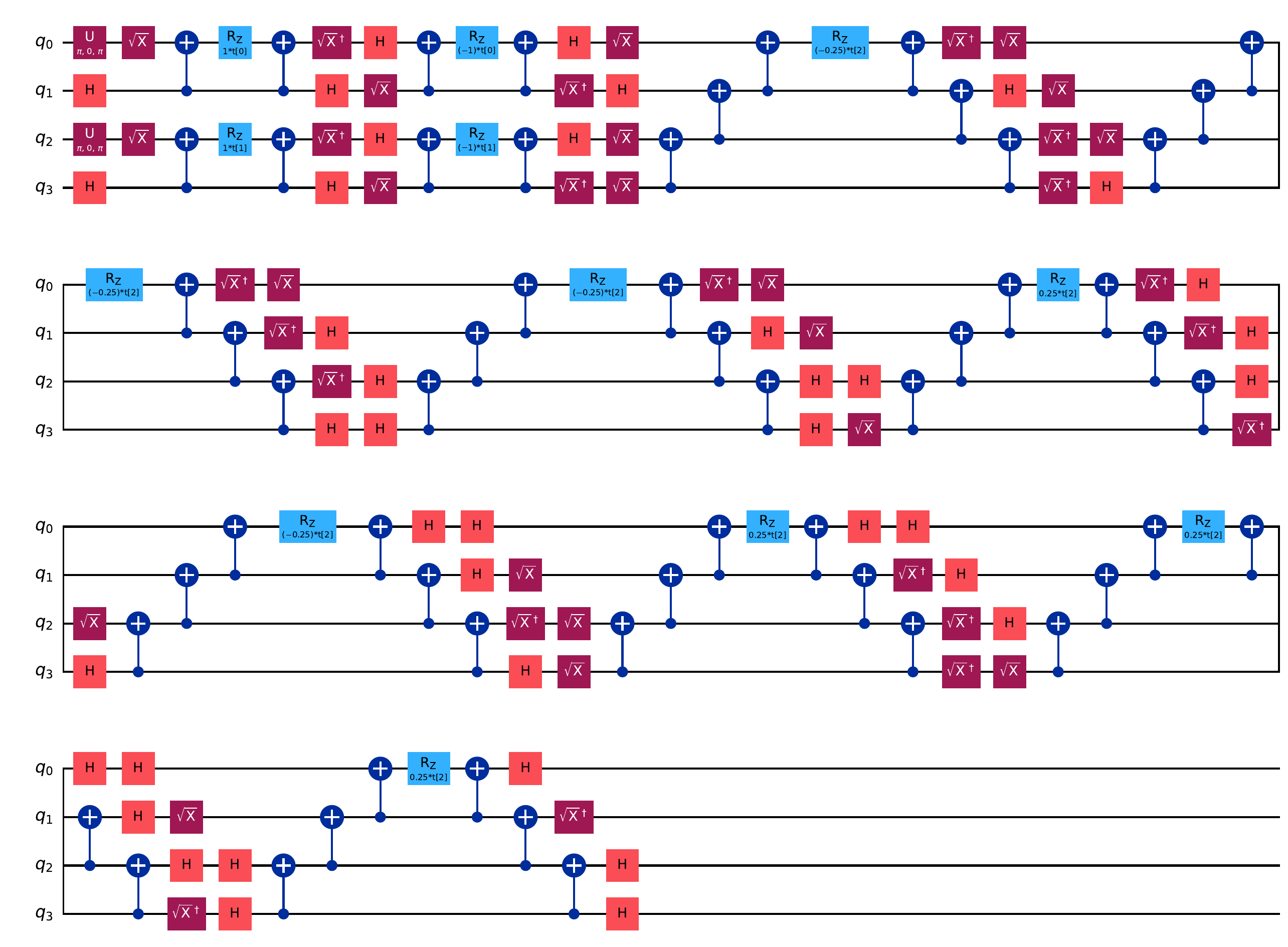}
    \caption{Qiskit circuit for the $H_2$ molecule decomposed to show the quantum gates. It comprises the Hartree-Fock state and the single and double excitation gates. The $U$ gates on $q_0$ and $q_2$ create the Hartree-Fock state. Parameter $t[0]$ of the $R_Z$ gate refers to the coefficient of $a^\dagger_1 a_0$, parameter $t[1]$ refers to the coefficient of $a^\dagger_3 a_2$, while parameter $t[2]$ is the coefficient of the double-excitation term $a^\dagger_1 a^\dagger_3 a_0 a_2$.}
    \label{fig:h2-2nd-dec}
\end{figure*}

Fig.~\ref{fig:h2-2nd-dec} shows the subsequent decomposition of these evolution operators into quantum gates. This construction follows the circuit-synthesis procedure described in Section~\ref{sec:circuit}. For each Pauli string, local basis transformations are first applied to map \(X\)- and \(Y\)-type operators onto the \(Z\)-basis. These transformations are implemented using gates such as the Hadamard (\(H\)) and \(R_x(\pi/2)\) gates, or equivalently the \(\sqrt{X}\) gate (see Appendix~\ref{subsec:sx}). Next, a ladder of CX gates computes the required qubit parity. The variational parameters \(t[0]\), \(t[1]\), and \(t[2]\) are then introduced through \(R_z\) rotations, whose angles are proportional to the corresponding coupled-cluster amplitudes. Finally, the parity-computation circuit and basis transformations are reversed, restoring the computational basis while preserving the desired unitary evolution.



\section{Acknowledgments}
This work was supported by the National Science and Technology Council (Taiwan) under grant NSTC 114-2112-M-182-002-MY3 and by Chang Gung Memorial Hospital under grant BMRPL94. 

\section{Code availability}
The Qiskit code used for analysis in this article can be accessed in GitHub at \url{https://github.com/Quantum-AI-Biomedical-Research-Lab/estimating-ground-state-energies}. 

The repository contains the following files: 
\begin{itemize}
    \item ``excitations-from-smiles.ipynb": user provides molecule name, the code extracts the molecule geometry and smiles code from PubChem, then uses PySCFDriver from Qiskit Nature to extract single and double excitations from UCCSD.   
    \item ``commutativity-checker.ipynb": commutativity checker for fermionic excitations in second quantization
    \item ``hamiltonian-su2-symmetry-checker.ipynb": checks if a given Hamiltonian is SU(2)-symmetric. 
    \item ``uccsd-vqe-ground-state-energy-h2.ipynb": VQE calculation of the ground state energy of $H_2$ using the UCCSD ansatz.
\end{itemize}



\bibliography{references}

\section*{Appendices}
\appendix

\section{Noncommutativity in UCCSD \label{app:A}}


The classical, non-unitary coupled-cluster singles and doubles (CCSD)
method defines the cluster operator as
\begin{equation}
T = T_1 + T_2,
\end{equation}
where
\begin{equation}
\begin{aligned}
T_1 &= \sum_{ia} t_i^a a_a^\dagger a_i,\\
T_2 &= \frac{1}{4}\sum_{ijab} t_{ij}^{ab}
a_a^\dagger a_b^\dagger a_j a_i.
\end{aligned}
\end{equation}
Here, $i,j$ denote occupied spin orbitals, while $a,b$ denote
virtual spin orbitals. The operators $T_1$ and $T_2$ describe
single and double excitations, respectively. The factor $1/4$
in $T_2$ accounts for the antisymmetry of the double-excitation
amplitudes and operators under the usual summation convention.

In conventional coupled-cluster theory, pure excitation operators
commute with one another. However, their products do not necessarily
vanish: compatible excitation operators can produce higher-order
excitations. Products involving repeated creation or annihilation
operators for the same fermionic mode, on the other hand, vanish
by the canonical anticommutation relations.

\subsection{An example of commuting excitation operators}

Consider the excitation operators
\begin{equation}
\begin{aligned}
E_i^a &= a_a^\dagger a_i,\\
E_{ij}^{ab} &= a_a^\dagger a_b^\dagger a_j a_i,
\end{aligned}
\end{equation}
where $a\ne b$ and $i\ne j$. Assume that $i,j$ label occupied
orbitals and $a,b$ label virtual orbitals, so that the occupied
and virtual indices are mutually distinct.

The commutator is
\begin{equation}
\begin{aligned}
[E_i^a,E_{ij}^{ab}]
={}&a_a^\dagger a_i a_a^\dagger a_b^\dagger a_j a_i\\
&-a_a^\dagger a_b^\dagger a_j a_i a_a^\dagger a_i.
\label{eq:com}
\end{aligned}
\end{equation}

Fermionic creation and annihilation operators satisfy the canonical
anticommutation relations:
\begin{equation}
\begin{aligned}
\{a_p,a_q\} &= 0,\\
\{a_p^\dagger,a_q^\dagger\} &= 0,\\
\{a_p^\dagger,a_q\} &= \delta_{pq},
\end{aligned}
\end{equation}
where $\{A,B\}=AB+BA$. Equivalently,
\begin{equation}
\begin{aligned}
a_p a_q &= -a_q a_p,\\
a_p^\dagger a_q^\dagger &= -a_q^\dagger a_p^\dagger,\\
a_p a_q^\dagger &= \delta_{pq}-a_q^\dagger a_p.
\end{aligned}
\end{equation}
In particular, setting the indices equal gives
\begin{equation}
a_p^2=(a_p^\dagger)^2=0.
\label{eq:nilpotent_fermionic}
\end{equation}

We can now simplify the two products in Eq.~(\ref{eq:com}).
For the first product,
\begin{equation}
\begin{aligned}
E_i^a E_{ij}^{ab}
&=a_a^\dagger a_i a_a^\dagger a_b^\dagger a_j a_i\\
&=-a_a^\dagger a_a^\dagger a_i a_b^\dagger a_j a_i\\
&=0,
\end{aligned}
\label{eq:left}
\end{equation}
where we used $a_i a_a^\dagger=-a_a^\dagger a_i$ and
$(a_a^\dagger)^2=0$.

For the second product,
\begin{equation}
\begin{aligned}
E_{ij}^{ab} E_i^a
&=a_a^\dagger a_b^\dagger a_j a_i a_a^\dagger a_i\\
&=-a_a^\dagger a_b^\dagger a_j a_a^\dagger a_i a_i\\
&=0,
\end{aligned}
\label{eq:right}
\end{equation}
where we used $a_i a_a^\dagger=-a_a^\dagger a_i$ and $a_i^2=0$.
Consequently,
\begin{equation}
[E_i^a,E_{ij}^{ab}]=0.
\end{equation}

In this example, both products vanish individually. 
The vanishing of a product is therefore not a general consequence of two operators sharing an orbital.

This example is directly applicable to the single excitaiton $a_1^\dagger a_0$ and the double excitation $a_1^\dagger a_3^\dagger a_0 a_2$ of the $H_2$ molecule.

\subsection{Noncommutativity in unitary coupled-cluster theory}

To demonstrate that the anti-Hermitian excitation operators of the previous example do not commute, consider
\begin{equation}
\begin{aligned}
\kappa_i^a
&=E_i^a-(E_i^a)^\dagger
=a_a^\dagger a_i-a_i^\dagger a_a,\\
\kappa_{ij}^{ab}
&=E_{ij}^{ab}-(E_{ij}^{ab})^\dagger\\
&=a_a^\dagger a_b^\dagger a_j a_i
-a_i^\dagger a_j^\dagger a_b a_a,
\end{aligned}
\end{equation}
where $a\ne b$ and $i\ne j$. 


Here, we can take advantage of the reference
Slater determinant:
\begin{equation}
|\Phi_0\rangle=a_i^\dagger a_j^\dagger|0\rangle,
\end{equation}
where $|0\rangle$ denotes the vacuum state.

Applying the double-excitation operator gives
\begin{equation}
\begin{aligned}
\kappa_{ij}^{ab}|\Phi_0\rangle
&=a_a^\dagger a_b^\dagger a_j a_i
a_i^\dagger a_j^\dagger|0\rangle\\
&=a_a^\dagger a_b^\dagger|0\rangle,
\end{aligned}
\end{equation}
where we used the canonical anticommutation relations and the
fact that the virtual orbitals are initially unoccupied.

Applying the single-excitation operator to this result yields
\begin{equation}
\begin{aligned}
\kappa_i^a\kappa_{ij}^{ab}|\Phi\rangle
&=-a_i^\dagger a_a a_a^\dagger
a_b^\dagger|0\rangle\\
&=-a_i^\dagger(1-a_a^\dagger a_a)
a_b^\dagger|0\rangle\\
&=-a_i^\dagger a_b^\dagger|0\rangle.
\end{aligned}
\end{equation}

In the reverse order, the single-excitation operator gives
\begin{equation}
\begin{aligned}
\kappa_i^a|\Phi\rangle
&=a_a^\dagger a_i a_i^\dagger
a_j^\dagger|0\rangle\\
&=a_a^\dagger a_j^\dagger|0\rangle.
\end{aligned}
\end{equation}
Since orbital $i$ is unoccupied in this resulting determinant
and orbital $b$ is also unoccupied, neither the double excitation
nor its de-excitation can act nontrivially. Therefore,
\begin{equation}
\kappa_{ij}^{ab}\kappa_i^a|\Phi\rangle=0.
\end{equation}

Combining the two results, we obtain
\begin{equation}
\begin{aligned}
[\kappa_i^a,\kappa_{ij}^{ab}]|\Phi\rangle
&=\kappa_i^a\kappa_{ij}^{ab}|\Phi\rangle
-\kappa_{ij}^{ab}\kappa_i^a|\Phi\rangle\\
&=-a_i^\dagger a_b^\dagger|0\rangle\\
&\ne 0.
\end{aligned}
\end{equation}
Consequently,
\begin{equation}
[\kappa_i^a,\kappa_{ij}^{ab}]\ne 0.
\end{equation}

Thus, although the pure excitation operators commute, their
anti-Hermitian combinations do not. 

\subsection{Implications for quantum circuit implementations}

The exact UCC ansatz is defined by the exponential of the sum of
its anti-Hermitian generators. If these generators do not commute,
the exponential generally cannot be replaced exactly by a product
of their individual exponentials. For example,
\begin{equation}
e^{A+B}\ne e^A e^B
\qquad\text{if }[A,B]\ne 0,
\end{equation}
in general. Quantum circuit implementations often approximate the
combined evolution using product formulas, such as the first-order
Trotter formula,
\begin{equation}
e^{A+B}
\approx
\left(e^{A/m}e^{B/m}\right)^m,
\end{equation}
where the approximation improves as the number of Trotter steps
$m$ increases under the usual convergence assumptions.

Consequently, noncommutativity matters when choosing the ordering
and grouping of operators in a product-formula implementation of
UCCSD. It is important to distinguish this circuit approximation
from the mathematical UCC ansatz itself. The ansatz is defined by
the full exponential, whereas the product formula is one possible
way to implement it approximately.

\section{The origin of $i$ in evolution operators} \label{app:i}

The reason why we want the operators in the exponent to be
anti-Hermitian rather than Hermitian is that we need unitary
operators to act on quantum states. In other words, the evolution
operator $U=e^A$ must be unitary, meaning that
\begin{equation}
U^\dagger U=UU^\dagger=I.
\end{equation}
This condition is satisfied if $A$ is anti-Hermitian,
$A^\dagger=-A$.

To see this, recall that
\begin{equation}
(e^A)^\dagger=e^{A^\dagger}.
\end{equation}
If $A$ is anti-Hermitian, then
\begin{equation}
(e^A)^\dagger=e^{-A}.
\end{equation}
Since $e^{-A}e^A=e^Ae^{-A}=I$, the operator $e^A$ is unitary.

In contrast, if $A$ is Hermitian, then
\begin{equation}
(e^A)^\dagger=e^A,
\end{equation}
so $e^A$ is Hermitian rather than generally unitary. In particular,
\begin{equation}
(e^A)^\dagger e^A=e^{2A}\ne I
\end{equation}
in general.


 

As for the time evolution operator $e^{iHt},$ we do not require $H$ to be Hermitian, because there is $i$ in the exponent which means 
\begin{equation}(e^{iHt})^{\dagger} = e^{{iHt}^{\dagger}} = e^{-iHt}\end{equation}

The time evolution operator arises from Schr\"odinger's equation
\begin{equation}i\hbar\partial_{t}\psi = H\psi\end{equation}

Consider the time-dependent Schr\"odinger equation
\begin{equation}i\frac{d}{dt}\psi(t) = H\psi(t)\end{equation} 

with $\hbar = 1$. We reorder the equation by moving $\psi$ to the left and $idt$ to the right to obtain 
\begin{equation}\frac{1}{\psi}d\psi = -iHdt\end{equation} 

where we omitted the time $t$ for simplicity. 

Then, we integrate both sides 
\begin{equation}\int \frac{1}{\psi}d\psi = \int -iHdt \Rightarrow \psi(t) = e^{-iHt}\psi(0)\end{equation}

where $ e^{-iHt}$ is the time evolution operator. 

\vspace{0.5cm}
\noindent \textit{Note: Why do the terms $e^A$ and $(e^A)^\dagger$ need to cancel out? Let's say $A$ in the the operator $e^A$ is Hermitian. Then, applying the operator $e^A$ to a quantum state $\ket{\psi}$ gives an evolved state:}
\begin{equation}e^A\ket{\psi} = \ket{\psi'}\end{equation}
\textit{The adjoint of this operation is}
\begin{equation}\bra{\psi'} = (e^A\ket{\psi})^\dagger = \bra{\psi}e^{A^\dagger} = \bra{\psi}e^{A}\end{equation}
\textit{Taking the inner product of $\ket{\psi'}$ with itself results in 1 as the vectors are assumed to be normalized:}
\begin{equation}\braket{\psi'}{\psi'} = 1\end{equation}
\textit{However, this is generally not the case for the evolution of the state $\ket{\psi}$:}
\begin{equation}\braket{\psi'}{\psi'} = \bra{\psi}e^{A}e^A\ket{\psi} = \bra{\psi}e^{2A}\ket{\psi} \not= 1\end{equation}

\section{Fermionic number operator}

The fermionic number operator for mode $p$ is defined as
\begin{equation}
n_p = a_p^\dagger a_p.
\end{equation}

Fermionic modes obey the Pauli exclusion principle, so the occupation number
\begin{equation}
n_p \in \{0,1\}.
\end{equation}

In quantum chemistry, each fermionic mode corresponds to a spin orbital.  
The number operator measures whether that orbital contains an electron.

The action of $n_p$ on the occupation basis states is:
\begin{itemize}
    \item If the state is empty:
\begin{equation}
a_p^\dagger a_p |0\rangle = a_p^\dagger (0) = 0.
\end{equation}

\item If the state is occupied:
\begin{equation}
a_p^\dagger a_p |1\rangle
= a_p^\dagger (a_p |1\rangle)
= a_p^\dagger |0\rangle
= |1\rangle.
\end{equation}
\end{itemize}

Thus, the operator measures whether the mode is empty or occupied.

Under the Jordan--Wigner transformation,
\begin{equation}
n_p = \frac{1}{2}(I - Z_p).
\end{equation}

Here is how to calculate it: 
\begin{equation}
a_p^\dagger a_p
=
\frac{X_p - iY_p}{2} \left( \prod_{j=0}^{p-1} Z_j \right)
\frac{X_p + iY_p}{2} \left( \prod_{j=0}^{p-1} Z_j \right)
\end{equation}
Since
\begin{equation}
\left( \prod_{j=0}^{p-1} Z_j \right)^2 = I,
\end{equation}
the $Z$-strings cancel:
\begin{equation}
a_p^\dagger a_p
=
\frac{1}{4}
(X_p - iY_p)(X_p + iY_p)
\end{equation}
Expand them:
\begin{equation}
\frac{1}{4}(X_p - iY_p)(X_p + iY_p)
=
\frac{1}{4}(X^2 + iXY - iYX + Y^2)
\end{equation}
In Pauli algebra it holds that:
\begin{equation*}
\begin{split}
X^2 &= Y^2 = I \\
XY &= iZ \\
YX &= -iZ
\end{split}
\end{equation*}
Therefore:
\begin{equation}
iXY = -Z,
\qquad
-iYX = -Z,
\end{equation}
Substituting into our calculation results in
\begin{equation}
\frac{1}{4}(2I - 2Z_p)
=\frac{1}{2}(I - Z_p)
\end{equation}

\end{document}